\documentclass[aps,preprintnumbers,superscriptaddress,nofootinbib,aps,prl,twocolumn,floatfix]{revtex4-1}
\pdfoutput=1

\usepackage{multirow,array}
\usepackage{amsmath,amssymb}
\usepackage{graphicx}
\usepackage{color}
\usepackage{url}
\usepackage{slashed}
\usepackage{multirow}
\usepackage{footmisc}
\usepackage{hyperref}
\usepackage{braket}
\usepackage{colortbl}
\usepackage{orcidlink}
\usepackage{physics}
\usepackage{mathrsfs}

\hypersetup{
  pdfauthor={Yang Ma and Zihui Wang}
  pdftitle={star cluster}
  pdfkeywords={star cluster}
}

\definecolor{nicered}{rgb}{0.5,0.,0.}
\definecolor{nicegreen}{rgb}{0.,0.5,0.}
\definecolor{niceblue}{rgb}{0.,0.,0.5}
\hypersetup{colorlinks,citecolor=nicegreen,linkcolor=nicered,urlcolor=niceblue}

\newcommand{\DM}{\mathrm{DM}}

\newcommand{\bmax}{b_{\mathrm{max}}}
\newcommand{\bmin}{b_{\mathrm{min}}}
\newcommand{\rmax}{r_{\mathrm{max}}}
\newcommand{\fdf}{F_{\mathrm{dyn}}}

\newcommand{\vs}{v_{\star}}
\newcommand{\sdm}{\sigma_{\DM}}
\renewcommand{\ss}{\sigma_{\star}}
\newcommand{\tre}{t_{\mathrm{relax}}}
\newcommand{\tage}{t_{\mathrm{age}}}

\newcommand{\ei}[1]{\mathrm{Ei}\qty(#1)}
\newcommand{\ta}{\tilde{\alpha}}

\begin{document}

\title{A new probe of dark matter-baryon interactions in compact stellar systems}
\author{Yang Ma\orcidlink{0000-0002-9419-6598}} 
\email{yang.ma@uclouvain.be}
\affiliation{INFN, Sezione di Bologna, via Irnerio 46, 40126 Bologna, Italy}
\affiliation{Center for Cosmology, Particle Physics and Phenomenology, Universit\'e catholique de Louvain, B-1348 Louvain-la-Neuve, Belgium}
\author{Zihui Wang\orcidlink{0000-0003-2407-1471}} 
\email{zihui.wang@nyu.edu}
\affiliation{Center for Cosmology and Particle Physics, Department of Physics, New York University, New York, NY 10003, USA}

\preprint{IRMP-CP3-25-02}

\begin{abstract}
 \noindent
We investigate the astrophysical consequences of an attractive long-range interaction between dark matter and baryonic matter. Our study highlights the role of this interaction in inducing dynamical friction between dark matter and stars, which can significantly influence the evolution of compact stellar systems. Using the star cluster in Eridanus II as a case study, we derive a new stringent upper bound on the interaction strength $\tilde{\alpha}\leq 314.5$ for the interaction range $\lambda = 1$ pc. This constraint is independent of the dark matter mass and can improve the existing model-independent limits on $\tilde{\alpha}$ by a few orders of magnitude. Furthermore, we observe that the constraint is insensitive to the mass of the stellar system and the dark matter density in the stellar system as long as the system is dark matter dominated. This new approach can be applied to many other stellar systems, and we obtain comparable constraints from compact stellar halos observed in ultrafaint dwarf galaxies.
\end{abstract}

\maketitle

\noindent
{\bf I. Introduction}\\
Exploring interactions between dark matter (DM) and baryons is crucial for understanding the nature of DM. 
There have been extensive studies on the possible observational consequences of a long-range interaction between DM and baryonic matter~\cite{Adelberger:2003zx, Fornengo:2011sz, Serebrov:2012vm, vandenAarssen:2012vpm, Gresham:2017zqi, Coskuner:2018are, Bhoonah:2020dzs, Xu:2020qjk, Gaidau:2021vyr, Gresham:2022biw, Cruz:2022otv, NANOGrav:2023hvm, Raj:2023azx}, inspired by various theoretical models of DM~\cite{Holdom:1985ag, Goldberg:1986nk, DeRujula:1989fe, Brahm:1989jh, Wise:2014jva, Wise:2014ola, Gresham:2017zqi, Coskuner:2018are, Farrar:2020zeo, Farrar:2023wta, Bhoonah:2020dzs, Gan:2023wnp, Detmold:2014qqa, Krnjaic:2014xza}.
In these models, the total interaction between a DM and baryonic object with gravity included can be described by the potential:
\begin{equation}
    V(r) = - \frac{G M_1 M_2}{r} (1+\ta e^{-r/\lambda}), \label{eq:YukawaPotential}
\end{equation}
where $G$ is Newton's constant, $\tilde{\alpha}$ is the strength of the long-range interaction relative to gravity, $\lambda$ characterizes the range of the interaction, and $M_{1,2}$ are masses of the DM and baryonic objects which can be fundamental particles or composite states. 
This type of interaction can emerge in models of millicharged DM~\cite{Holdom:1985ag, Goldberg:1986nk, DeRujula:1989fe, Brahm:1989jh, Gan:2023jbs}, or more generally in models where baryons and DM couple to a scalar or vector mediator~\cite{Wise:2014jva, Wise:2014ola, Gresham:2017zqi, Coskuner:2018are, Farrar:2020zeo, Farrar:2023wta, Bhoonah:2020dzs, Gan:2023wnp}. It can also arise in a different class of DM models where confinement within the dark gauge group produces dark atoms and dark nuclei~\cite{Detmold:2014qqa, Krnjaic:2014xza}.

At distances $\lesssim \lambda$, Eq.~\eqref{eq:YukawaPotential} describes the effective gravity between DM and baryons. Since the inverse-square law of gravity is tested with high precision, the value of $\ta$ must not be relatively large~\cite{Adelberger:2003zx}. Additionally, upper bounds on $\ta$ have been established from pulsar timing arrays~\cite{NANOGrav:2023hvm}, neutron star heating~\cite{Gresham:2022biw}, superbursts from neutron stars and white dwarfs~\cite{Raj:2023azx}, and dark matter-baryon scattering measurements~\cite{Xu:2020qjk}. In this letter, we suggest a new method to probe the long-range interaction between DM and baryons in  Eq.~\eqref{eq:YukawaPotential} by studying its impact on compact stellar systems and derive constraints on $\ta$. These constraints are  considered ``model-independent'' as they do not rely on the specific microphysics of DM beyond the assumption of Eq.~\eqref{eq:YukawaPotential}. 

Ultrafaint dwarf galaxies (UFDs) are ideal astrophysical targets to detect signals of DM interactions~\cite{WadFar21,LahLu20,Kim20,TakLu21, Wadekar:2021qae, Wadekar:2022ymq, Graham:2023unf, Graham:2024hah, Colafrancesco:2006he, Baltz:2008wd, Caputo:2018ljp, Regis:2020fhw}.
Eridanus II (Eri II) is presently the least luminous UFD observed to host a globular cluster~\cite{Koposov:2015cua, Bechtol_2015, Crnojevic_2016, DES:2016vji,Contenta, Simon:2020qsf}. Located near the center of Eri II, the star cluster offers unique opportunities to probe the density profile of DM~\cite{Amorisco:2017hac,Contenta,Simon:2020qsf, Orkney}. 
This star cluster has also been used to constrain various DM models such as primordial black hole (PBH) and ultralight DM as these DM candidates could heat the star cluster causing the cluster to expand over time~\cite{Brandt:2016aco, Zoutendijk2020,Koulen:2024emg, Marsh:2018zyw}. 
In this letter we will demonstrate that the star cluster may also contract if it experiences dynamical cooling from DM, providing another avenue for studying DM.

In a series of seminal work~\cite{chand43,chandrasekhar1943dynamical2,chandrasekhar1943dynamical3}, Chandrasekhar studied the frictional force experienced by a subject star in a background of field stars resulting from cumulative gravitational scattering, i.e., dynamical friction. Similarly, as a star moves in a DM halo, the net interaction between the star and DM particles exerts dynamical friction on the star.
The presence of a long-range interaction given by Eq.~\eqref{eq:YukawaPotential} can induce additional dynamical friction between stars and DM apart from the standard gravitational contribution.
If DM is lighter than stars, dynamical friction leads to the loss of stars' kinetic energy, driving the stellar system to contract. 

Model-independent limits on $\ta$ can then be determined by studying the shrinking of the stellar system's half-light radius.
For the star cluster in Eri II, our method covers a wide range of DM masses and results in upper bounds $\ta\leq \order{100}$. These results provide valuable complementary constraints on the DM-baryon long-range interaction, alongside the existing ones from Refs.~\cite{NANOGrav:2023hvm, Xu:2020qjk, Raj:2023azx, Gresham:2022biw}.
In addition to the star cluster in Eri II, our method can be broadly applied to many other compact stellar systems.

\vskip 0.1cm
\noindent
{\bf II. Evolution of stellar systems}\\
Chandrasekhar's original formula for dynamical friction can be derived by various methods, such as perturbation theory of Kepler orbits~\cite{chand43}, Fokker-Planck diffusion coefficients~\cite{Binney08} or the kinetic energy gain of field stars~\cite{Binney08,Aceves06}. 
The derivation using kinetic energy gain can be naturally extended to deduce dynamical friction induced by the long-range potential in Eq.~\eqref{eq:YukawaPotential} between stars and DM particles. 
We focus on the case where stars lose kinetic energy due to dynamical friction, which occurs when stars are heavier than DM. The cooling rate of a star can be written as 
\begin{equation}
    \fdf(\ta,\lambda) \vs = -\frac{4\pi G^2 m_{\star}^2\rho_{\DM}}{\vs} \mathcal{G}(X) \mathcal{I}(\ta,\lambda),
    \label{eq:df1}
\end{equation}
where we have assumed a uniform DM density $\rho_{\DM}$, $m_{\star}$ is the mass of a single star, $\vs$ is the star's velocity, and the function $\mathcal{G}(X)$ gives the fraction of DM particles with velocity $\leq \vs$
\begin{eqnarray}
    \mathcal{G}(X) = \erf(X) - \frac{2X}{\sqrt{\pi}} e^{-X^2}, ~~~X=\frac{\vs}{\sqrt{2}\sdm},
\end{eqnarray}
with $\sdm$ being the Maxwellian velocity dispersion of DM.
The form factor $\mathcal{I}(\ta,\lambda)$ arises from integrating the kinetic energy gain of DM particles encountered by the subject star over the impact parameter $b$
\begin{eqnarray*}
    &&\mathcal{I} (\ta,\lambda) = \int_{\bmin}^{\bmax} \frac{(1+\ta (1+b/\lambda) e^{-b/\lambda})^2}{b}{\rm d}b \,  \nonumber\\
    &=&\left.\left[\ta^2\ei{\frac{-2b}{\lambda}} + 2\ta\ei{\frac{-b}{\lambda}} + \ln b -{\mathscr E}(b/\lambda)\right]\right|^{\bmax}_{\bmin},
    \label{eq:CoulombLog2}
\end{eqnarray*}
where $\ei{x}\equiv -\int^\infty_x dt\, e^{-t}/t$ and ${\mathscr E}(x)=2 \ta e^{-x}+ (5\ta^2/4+\ta^2 x/2)e^{-2x}$.
In practice, $\bmin = Gm_{\star}/\sdm^2$ and $\bmax$ is the size of the astrophysical system~\cite{Brandt:2016aco}.
Note that Eq.~\eqref{eq:df1} only accounts for the first-order diffusion coefficient $D[\Delta v_{\parallel}]$, which is sufficient when DM is much lighter than stars. In the limit $\ta\to 0$ or $\lambda \to 0$, $\mathcal{I} (\ta,\lambda)$ becomes the well-known Coulomb logarithm $\ln \Lambda \equiv \ln(\bmax/\bmin)$ and Eq.~\eqref{eq:df1} restores Chandrasekhar's formula of gravity-induced dynamical friction. 

The potential energy of a star is also influenced by the long-range interaction with DM. If a stellar system is embedded in a background of DM particles with uniform density $\rho_{\DM}$, the potential energy per unit mass of the system is
\begin{equation}
    \frac{U}{M_{\star}} = -\beta_1 \frac{GM_{\star}}{r_h} + \beta_2 G\rho_{\DM} r_h^2 + u(\ta, \lambda, r_h) + \,\mathrm{const.},
\label{eq:PE}
\end{equation}
where $M_{\star}$ is the stellar mass of the system, $r_h$ is its half-light radius, and $\beta_{1,2}$ are constants determined by the density profile of the stellar system.
In Eq.~\eqref{eq:PE}, the first two terms represent the potential energy from the self-gravitation of the system and the gravitational field of DM~\cite{Brandt:2016aco}, and the third term $u(\ta, \lambda, r_h)$ arises from the long-range interaction between DM and stars. 

Following the method of Ref.~\cite{Brandt:2016aco}, the evolution of $r_h(t)$ can be obtained by differentiating Eq.~\eqref{eq:PE} with respect to time. Combining this with the virial theorem 
and considering the velocity distribution of $\vs$,
we find
\begin{equation}
\dv{r_h}{t} = \frac{2M_{\star}}{m_{\star}} \qty(\frac{\partial U}{ \partial r_h } )^{-1} \int d\vs\, f(\vs) \fdf \vs,
\label{eq:rh}
\end{equation}
where the function $f(\vs)$ is the Maxwellian velocity distribution of stars with dispersion $\ss$. Notably, Eq.~\eqref{eq:rh} is independent of the DM mass $m_{\DM}$ as long as $m_{\DM} \ll m_{\star}$.\footnote{If $m_{\DM}\gtrsim m_{\star}$, higher order diffusion coefficients proportional to $m_{\DM}/m_{\star}$ need to be included to model dynamical friction~\cite{Binney08}. It should also be noted that Eq.~\eqref{eq:rh} might not directly apply to ultralight DM with $m_{\DM} \sim 10^{-22}$~eV. Fluctuations at the de Broglie scale of ultralight DM can produce granules with masses $\gtrsim \order{1000\, M_\odot}$~\cite{Schive:2014dra, Hui:2016ltb, Veltmaat:2018dfz, Hui:2021tkt} which can dynamically heat stars~\cite{Marsh:2018zyw, Dalal:2022rmp, Wadekar:2022ymq}.}

In order to determine the time evolution of $r_h(t)$, the form of $\partial u(\ta, \lambda, r_h)/\partial r_h$ is needed. In the $\lambda \to \infty$ limit, Eq.~\eqref{eq:YukawaPotential} becomes 
\begin{eqnarray}
    \lim_{\lambda\to \infty} V(r) = - (1+\ta) G M_1 M_2/r,
\end{eqnarray}
so that $u(\ta, \lambda, r_h) \to \beta_2 \ta G\rho_{\DM} r_h^2 $.
In the regime $\bmin \lesssim \lambda \lesssim r_h$, we demonstrate that $\partial u(\ta, \lambda, r_h)/\partial r_h=0$.
The function $u(\ta, \lambda, r_h)$ in Eq.~\eqref{eq:PE} can be obtained by integrating 
a single star's potential energy $u_{\star}(r)$
\begin{equation}
    u = \frac{\int_0^{r_h} d^3r\, u_{\star}(r)n_{\star}(r)}{M_{\star}} = \frac{\int_0^{r_h} d^3r\, u_{\star}(r) n_{\star}(r)}{m_{\star} \int_0^{r_h} d^3r\, n_{\star}(r)},
    \label{eq:u}
\end{equation}
where $r$ is the radial coordinate of the star in the system and $n_{\star}(r)$ is the number density of stars.
Since the long-range interaction is exponentially suppressed outside the interaction range $\lambda$, $u_{\star}$ is dominantly contributed by DM particles within a distance of $\order{\lambda}$ from the star. Therefore, $u_{\star}$ is independent of the radial coordinate $r$ of the star and consequently, $\partial u/\partial r_{h}=0$.\footnote{This result is derived under the assumption that the stellar system resides in a DM core. If the DM follows a cuspy Navarro–Frenk–White  distribution rather than a cored one, we estimate that $\partial u/\partial r_h \propto e^{-r_h/\lambda}$ for a uniform stellar profile or an exponential Sersic profile $n_{\star}(r) \sim e^{-r/r_h}$. Consequently, for $\lambda \ll r_h$, the partial derivative $\partial u/\partial r_h$ remains significantly suppressed and Eq.~\eqref{eq: du} still continues to provide a good approximation.} This is distinct from the gravitational case, where the gravitational force on the star is provided by DM within a distance of $r$ from the center of the stellar system. In summary, the partial derivative of Eq.~\eqref{eq:PE} takes the following form
\begin{eqnarray}
\frac{\partial U}{M_{\star}\partial r_h} \simeq
\begin{cases}
    \beta_1 \frac{GM_{\star}}{r_h^2} + 2\beta_2 G\rho_{\DM} r_h ,\quad & 0< \lambda \lesssim r_h\\
    \beta_1 \frac{GM_{\star}}{r_h^2} + 2\beta_2 (1+\ta) G\rho_{\DM} r_h ,\quad & \lambda \gg r_h .
\end{cases}  
\label{eq: du}
\end{eqnarray}
With Eq.~\eqref{eq:df1} and~\eqref{eq: du}, the differential equation in Eq.~\eqref{eq:rh} can be solved to obtain the evolution of $r_h(t)$.

It is worth noting that in Eq.~\eqref{eq:rh} the weak encountering between stars, i.e., the two-body relaxation, is not included. The velocity fluctuation from cumulative weak encountering can push stars to diffuse into the outer part and critically impact the evolution of the stellar system, which typically occurs after a few relaxation time scales~\cite{Binney08}. The relaxation time scale $\tre$ is defined to be the time required for the root-mean-square velocity fluctuation to equal the original velocity and can be found by~\cite{Binney08}
\begin{equation}
    \tre = \frac{\vs^3}{8\pi G m_{\star}^2 n_{\star} \ln\Lambda},
\label{eq:trelax}
\end{equation}
where $n_{\star}$ is the number density of stars and $\ln\Lambda$ is the Coulomb logarithm of the stellar system. To minimize the potential impact of relaxation, it is advisable to model the evolution of $r_h(t)$ by Eq.~\eqref{eq:rh} over a time interval shorter than $\tre$.

\vskip 0.1cm
\noindent
{\bf III. Probing dark matter in Eridanus II}\\
As mentioned above, the star cluster in Eri II has been recognized as a powerful DM tracer.
At a distance of $\sim 366$ kpc from the Milky Way, Eri II is a UFD with absolute magnitude $M_V = -7.1\pm 0.3$, a mass-to-light ratio of $420^{+210}_{-140}\, M_\odot/L_\odot$, and a compact star cluster located in its central region~\cite{Koposov:2015cua, Bechtol_2015, Crnojevic_2016, DES:2016vji,Contenta, Simon:2020qsf,Zoutendijk2020}. Observations of Eri II favor a cored DM profile~\cite{Contenta, Orkney}. 
The measured properties of Eri II and the star cluster reported by Refs.~\cite{Crnojevic_2016, Zoutendijk2020} are summarized in Table~\ref{tab:para}.
\begin{table}[htb]
\begin{ruledtabular}
    \begin{tabular}{l c c c}
    Parameter  & \cite{Crnojevic_2016} & \cite{Zoutendijk2020} \\ \hline
    Eri II $R_h$ (pc) & $277\pm 14$ &  \\
    Cluster $r_h$ (pc) & $13\pm 1$ &  \\
    $M_{\star}$ ($M_\odot$) & [2000, 6000] &  \\
    Cluster age $\tage$ (Gyr) & [3, 12] & \\
    $\sdm^{\mathrm{1D}}$ (km/s) &  & $10.3^{+3.0}_{-3.2}$ \\
    $\ss^{\mathrm{1D}}$ (km/s) &  & $2.3^{+5.3}_{-2.3}$ \\
    $\rho_{\DM}$ ($M_\odot/\mathrm{pc}^3$) &  & $7.1^{+6.4}_{-3.7}$ \\
    \end{tabular}
    \end{ruledtabular}
    \caption{Summary of the measured properties of Eri II and its central star cluster. From top to bottom are the half-light radius of Eri II, the half-light radius of the star cluster, the stellar mass of the cluster, the age of the cluster, the one-dimensional (1D) velocity dispersion of DM, the 1D velocity dispersion of the star cluster, and the DM density. 
       }
    \label{tab:para}
\end{table}

\begin{figure}[htb]
    \includegraphics[width=0.46\textwidth]{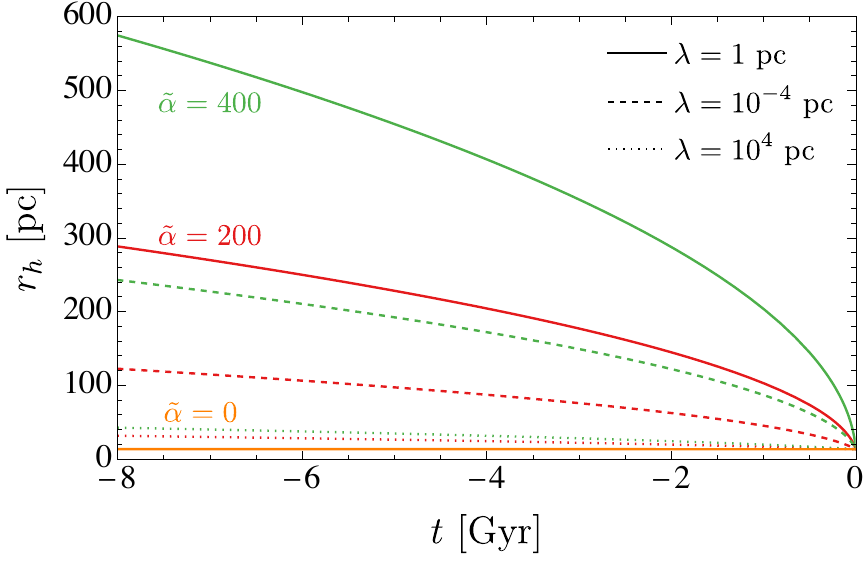}
    \caption{Time evolution of $r_h(t)$ for $\ta=400$ (green) and 200 (red), and $\lambda = 10^{-4}$ pc (dashed), 1 pc (solid), $10^4$ pc (dotted). The gravity-only case $\ta=0$ is also shown as a reference by the orange line.
    }
    \label{fig:rh}
\end{figure}

The time evolution of this star cluster's half-light radius $r_h(t)$ is described by Eq.~\eqref{eq:rh}. With the parameters taken from Table~\ref{tab:para}, we solve the differential equation for $t\leq 0$ to infer its $r_h$ in the past. Note that the three-dimensional velocity dispersion entering Eq.~\eqref{eq:rh} are given by $\sdm = \sqrt{3} \sdm^{\mathrm{1D}}$ and $\sigma_{\star}= \sqrt{3} \ss^{\mathrm{1D}} $. As a representative example we pick $M_{\star} = 3000\, M_\odot$ which corresponds to an age of 8 Gyr~\cite{Crnojevic_2016}. The boundary condition is set as $r_h(0)=13$ pc and conservatively $\bmax$ is also chosen to be $\bmax = r_h(0)$. With this setup, $r_h(t)$ is illustrated in Fig.~\ref{fig:rh} for various values of $\ta$ and $\lambda$. We notice that the evolution shows no dependence of $\rho_{\DM}$ and $M_{\star}$ as long as the star cluster is DM dominated, which agrees with the observation in Ref.~\cite{Brandt:2016aco}.

Three values of $\lambda$ are displayed in Fig.~\ref{fig:rh}: $\lambda = 10^{-4}$ pc (close to $\bmin$), $\lambda=1$ pc (between $\bmin$ and $\bmax$), and $\lambda=10^4$ pc (much larger than $\bmax$). 
As presented in Fig.~\ref{fig:rh}, $r_h$ evolves differently in the above three regimes of $\lambda$. For $\bmin \lesssim \lambda \lesssim \bmax$, the evolution of $r_h$ strongly depends on the value of $\ta$.
In the other two regimes, $\lambda \lesssim \bmin$ and $\lambda \gg \bmax$, the evolution of $r_h(t)$ rapidly converges to the gravity-only case $\ta = 0$. This is because in these regimes, the long-range force is either negligible or indistinguishable from gravity.  

Since any astrophysical system must have a finite size, $r_h$ in the past cannot be arbitrarily large. This requirement can be used to place constraints on $\ta$ as the time evolution of $r_h(t)$ derived above is highly sensitive to $\ta$ for $\bmin\lesssim\lambda\lesssim\bmax$.
More concretely, we impose an upper bound on $r_h(t)$ at a checkpoint $t = -t_0$ of special interest, i.e., 
\begin{equation}
    r_h(-t_0)\leq r_{\mathrm{max}}.
\label{eq:rmax}
\end{equation}
A natural choice of $t_0$ and $r_{\mathrm{max}}$ is the age of the cluster and the half-light radius of the galaxy $R_h$, which ensures the star cluster always remains in the bounds of the galaxy over the given timescale. 
To obtain limits on $\ta$, $r_{\mathrm{max}}$ is taken to be the central value of $R_h=277$ pc, and $t_0$ is set by the minimum value of $\tage=3$ Gyr which corresponds to $M_{\star}=2000\, M_\odot$,  
\begin{equation}
\begin{aligned}
    &\ta \leq 314.5, \qq{for} \lambda = 1\,\mathrm{pc}, \\
    &\ta \leq 747.1, \qq{for} \lambda = 10^{-4}\,\mathrm{pc}.
\label{eq:limits}
\end{aligned}
\end{equation}

\begin{figure}[thb]
    \includegraphics[width=0.46\textwidth]{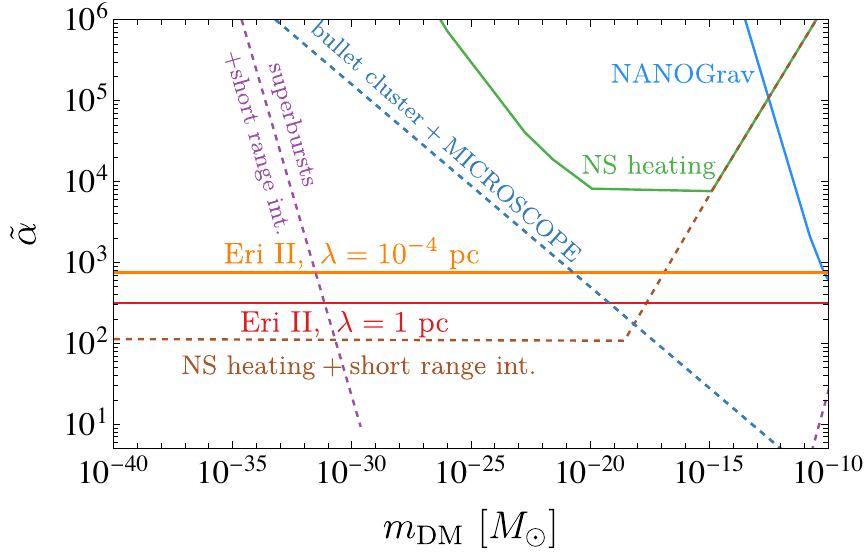}
    \caption{Compilation of upper limits on $\ta$. The orange and red lines are derived in this work based on observed properties of the star cluster in Eri II for $\lambda = 10^{-4}$ pc and 1 pc respectively. Other lines in the figure are from pulsar timing array observations by NANOGrav~\cite{NANOGrav:2023hvm} (azure) and neutron star heating~\cite{Gresham:2022biw} (green). Model-dependent limits, which rely on additional assumptions, are shown by dashed lines. These include superbursts of white dwarfs and neutron stars~\cite{Raj:2023azx} (dashed violet), maximal neutron star heating~\cite{Gresham:2022biw} (dashed brown) and combined results from the bullet cluster and MICROSCOPE~\cite{Gresham:2022biw} (dashed blue). }
    \label{fig:constraints}
\end{figure}

In deriving Eq.~\eqref{eq:limits}, no additional assumptions on the DM model besides Eq.~\eqref{eq:YukawaPotential} are employed, making the constraint model-independent. 
Furthermore, as noted previously, the time evolution of $r_h(t)$ is independent of the DM mass and density, and so is the constraint in Eq.~\eqref{eq:limits}. 
We plot Eq.~\eqref{eq:limits} by the solid red and orange lines in Fig.~\ref{fig:constraints}, alongside various existing limits.\footnote{Since photometric microlensing surveys exclude objects heavier than $\order{10^{-10}\, M_\odot}$ to wholly constitute the entire DM density~\cite{MACHO:1998qtf,Niikura:2017zjd, Croon:2020wpr,Croon:2020ouk}, we do not show constraints above this mass in Fig.~\ref{fig:constraints}.}
The azure and green solid lines are existing model independent constraints with $\lambda = 1$ pc from the 15-year dataset of pulsar timing array observations by NANOGrav~\cite{NANOGrav:2023hvm} and neutron star (NS) heating~\cite{Gresham:2022biw}.
The new constraint based on the time evolution $r_h(t)$ of the star cluster in Eri II could improve the existing model independent limits on $\ta$ in a broad range of DM masses.
For further reference, we also show limits on $\ta$ arising from superbursts of NS and white dwarfs~\cite{Raj:2023azx} (dashed violet), maximal NS heating~\cite{Gresham:2022biw} (dashed brown) and the combined results from the bullet cluster and MICROSCOPE~\cite{Gresham:2022biw} (dashed blue), which rely on additional assumptions on the DM model such as the existence of a short-range interaction between DM and baryons.

We comment on the possibility that only a fraction, $f_{\DM}$, of DM interacts with baryons through the long-range interaction in Eq.~\eqref{eq:YukawaPotential}. In this case, the right-hand side of Eq.~\eqref{eq:df1} and the function $u$ in Eq.~\eqref{eq:PE} need to be multiplied by $f_{\DM}$. We observe that the upper bound on $\ta$ from Eri II weakens approximately by a factor of $1/f_{\DM}$. 

\vskip 0.1cm
\noindent
{\bf IV. Discussion and conclusions}\\
As demonstrated with the star cluster in Eri II, our method produces stringent constraints on the long-range interaction between dark matter and baryons among the existing model-independent ones. Additionally, we observed that the constraints are independent of the DM mass, $m_{\rm DM}$, and its dependence on the DM density at the location of the stellar system, $\rho_{\rm DM}$, and the stellar system's mass, $M_{\star}$, are negligible. A few comments are made as following:
\begin{itemize}
    \item In our demonstration, we have chosen the lower limit on the star cluster's age, 3 Gyr, as $t_0$. It is noticed that by requiring $r_h(-t_0)\leq r_{\mathrm{max}}$, a larger value of $t_0$ manifestly leads to a stronger constraint on $\ta$. 
    Choosing $t_0=\tre=5.9$ Gyr could produce a stronger constraint 
    $\ta\leq 224.0$ for $\lambda = 1$ pc.

    \item The choice of $r_{\rm max}$ also affects the constraint on $\ta$. The benchmark constraints in Eq.~\eqref{eq:limits} are given by setting $\rmax$ to be the central value of $R_h = 277$~pc. To understand the $\rmax$ dependence, we observe that upon replacing $r_{\rm max}= R_h$ by $x R_h$ with $x \in [1/4,4]$, the upper limits in Eq.~\eqref{eq:limits} scale proportionally by a factor of $x$. We note that smaller $\rmax$ may be physically motivated. For instance, the maximum initial $r_h$ of the star cluster in Eri II is taken to be less than 20 pc in simulations~\cite{Contenta}. If we set $\rmax = 20$ pc, we can obtain much tighter limits.
 
    \item We further assess how the constraint on $\ta$ depends on the velocity dispersion $\sdm$ and $\ss$. As can be seen in Eq.~\eqref{eq:df1} and~\eqref{eq:rh}, a larger $\sigma_{\rm DM}$ leads to a weaker constraint, while a larger $\ss$ results in a stronger constraint. To illustrate the impact of these two parameters, we vary $\sigma_{\rm DM}$ and $\sigma_{\star}$ by a factor of 1/2 (2) and with $\lambda = 1$ pc, we find 
    $\ta \leq 613.1$ (173.3) for variations in $\ss$ and 
    $\ta \leq 130.7$ (818.2) for variations in $\sigma_{\rm DM}$.
    
    \item Our constraints on $\ta$ can be corroborated by the stability of the stellar system. A mild stability constraint requires that $r_h$ does not change by a factor of two within one crossing time $t_c \sim r_h/\ss$. This condition leads to 
    $\ta \leq 786.4$ for $\lambda = 1$ pc, in broad agreement with our results. 
\end{itemize}

Our modeling of the star cluster in Eri II are based on the assumption of a cored DM profile, which is favored by observations of Eri II~\cite{Contenta,Orkney}. However, we have demonstrated that our methodology can be reasonably adapted for a cuspy profile. Consequently, our approach can be applied to various stellar systems in other galaxies.
For example, we can require the half-light radius of the compact stellar halo in an ultrafaint dwarf galaxy (UFD) not to exceed that of the DM halo, allowing for the derivation of independent constraints on $\ta$. Many UFDs are observed to have compact stellar halos with a half-light radius $r_h \sim 30$ pc, a stellar mass $\sim 3000\, M_\odot$, an age $\tage \sim 10$ Gyr, a DM density $\rho_{\DM}\sim 1\, M_\odot/\mathrm{pc}^3$ and velocity dispersion $\sdm\sim 15$ km/s and $\ss\sim 5$ km/s~\cite{Martin2007, Geha2009, Belokurov2009, Simon2011, Koposov:2015cua, Bechtol_2015, Laevens2015, Brandt:2016aco}, with examples including Willman I, Segue I, Segue II, among others. 
The scale radius of the DM halo in UFDs typically ranges from 100 to 1000 pc (see e.g.~\cite{Fermi-LAT:2010cni,MAGIC:2011nta, VERITAS:2012cdo, MAGIC:2016xys}), and for a conservative estimate, we adopt 1000 pc as $\rmax$. By imposing $r_h(-\tage) \leq \rmax$, we find the upper limit 
$\ta \leq 405.5$ for $\lambda = 1$ pc, which is comparable to the constraint derived from the star cluster in Eri II. Upcoming surveys (e.g.~\cite{Drl19_LSST, Mut21}) are expected to discover a much larger number of UFDs, enabling broader application of our method. 

In conclusion, we have introduced a novel method to probe long-range DM-baryon interactions by leveraging the observed half-light radius of compact stellar systems. Our approach utilizes the drag force experienced by stars due to the long-range interaction with DM. The result provides a valuable complement to existing constraints on the long-range interaction~\cite{NANOGrav:2023hvm, Xu:2020qjk, Raj:2023azx, Gresham:2022biw}.

\vskip 0.1cm
\textbf{Acknowledgment:}
We thank Andrei Gruzinov, Nirmal Raj, Huangyu Xiao, Shengqi Yang, and Zhichao Zeng for useful discussions. YM is a Postdoctoral Fellow of the Fond de la Recherche Scientifique de Belgique (F.R.S.-FNRS), Belgium. The work of ZW was supported in part by a James Arthur Fellowship. 
  

\bibliographystyle{apsrev}
\bibliography{ref.bib}

\end{document}